\documentclass{Interspeech2024}
\usepackage{amssymb}
\usepackage{longtable}
\usepackage{multirow}
\usepackage{graphicx}

\interspeechcameraready 

\title{RaD-Net 2: A causal two-stage repairing and denoising speech enhancement network with knowledge distillation and complex axial self-attention\vspace{-18pt}}

\name[affiliation={1,2}]{Mingshuai}{Liu}
\name[affiliation={2}]{Zhuangqi}{Chen}
\name[affiliation={1}]{Xiaopeng}{Yan}
\name[affiliation={1}]{Yuanjun}{Lv}
\name[affiliation={2}]{Xianjun}{Xia} 
\name[affiliation={2}]{\\Chuanzeng}{Huang}
\name[affiliation={2}]{Yijian}{Xiao}
\name[affiliation={1*}]{Lei}{Xie}

\address{
  $^1$Audio, Speech and Language Processing Group (ASLP@NPU), School of Software, \\ Northwestern Polytechnical University, Xi'an, China\\
  $^2$ByteDance, China}
\email{liumingshuai@mail.nwpu.edu.cn, lxie@nwpu.edu.cn\thanks{* Corresponding author.}}


\begin{document}

\maketitle

\begin{abstract}
    
In real-time speech communication systems, speech signals are often degraded by multiple distortions. Recently, a two-stage Repair-and-Denoising network (RaD-Net) was proposed with superior speech quality improvement in the ICASSP 2024 Speech Signal Improvement (SSI) Challenge. However, failure to use future information and constraint receptive field of convolution layers limit the system's performance. To mitigate these problems, we extend RaD-Net to its upgraded version, RaD-Net 2. Specifically, a causality-based knowledge distillation is introduced in the first stage to use future information in a causal way. We use the non-causal repairing network as the teacher to improve the performance of the causal repairing network. In addition, in the second stage, complex axial self-attention is applied in the denoising network's complex feature encoder/decoder. Experimental results on the ICASSP 2024 SSI Challenge blind test set show that RaD-Net 2 brings 0.10 OVRL DNSMOS improvement compared to RaD-Net.
\end{abstract}

\section{Introduction}

Real-time speech communication systems play a crucial role in daily life and work. However, speech signals may suffer from multiple distortions during speech communication due to various factors, such as hardware limitations and network quality. These distortions include coloration, discontinuity, loudness, noisiness, and reverberation, severely damaging speech signal quality. Although researchers have invested considerable efforts, there is still a long way to go to make communication systems as good as or better than face-to-face communication. 

Over the past few decades, researchers have proposed many classical speech enhancement (SE) methods to remove noise from noisy signals. These methods include codebook-based approaches~\cite{codebook}, Wiener filtering-based strategies~\cite{wiener}, and nonnegative matrix factorization (NMF) techniques~\cite{nmf1}. In recent years, with the development of deep neural networks (DNNs), DNN-based methods have been widely used in the SE field and achieved superior performance compared to classical methods~\cite{conv-tasnet, dccrn}. However, these DNN-based methods often induce additional distortions to the speech while suppressing noise. Recently, a two-stage strategy has been proposed to address this issue~\cite{two-stage-1_1, two-stage-1_2, two-stage-1_3}. Specifically, the first-stage module is adopted for noise suppression. Then, in the second stage, the speech enhanced by the first stage is restored to higher-quality speech signals. With advancements in DNN-based SE techniques, researchers pay more attention to addressing various distortions that affect speech components in the real world. The ICASSP SSI Challenge Series~\cite{sig2023, sig2024} focuses on improving speech signal quality in communication systems using causal models. In these challenges, a ``restoration and enhancement" framework has been proposed to restore speech that is heavily degraded due to complex acoustic environments~\cite{gesper, ssi-net, ks-net, liu2023}. These methods apply two-stage processing modules. In the first stage, the restoration module processes frequency response distortions, isolated and non-stationary distortions, and loudness issues. Meanwhile, preliminary denoising and dereverberation are performed in this stage. Subsequently, the second stage module is adapted to remove residual noise and artifacts. 

Following the two-stage framework mentioned above, RaD-Net~\cite{rad-net} achieved superior performance in the ICASSP 2024 SSI Challenge~\cite{sig2024}. Specifically, a complex mapping-based model named COM-Net is applied in the repairing stage, while a complex masked-based model called S-DCCSN is adapted in the denoising stage. Furthermore, adversarial training is introduced to improve speech naturalness. However, the performance of the repairing network is limited without accessing future information. In addition, due to the constrained receptive field of convolution layers~\cite{frcrn}, the denoising network module's complex feature encoder/decoder could not effectively capture long-range relations between features.

To address the above limitations, we extend RaD-Net to RaD-Net 2, using different optimization strategies for each stage to further improve perceived speech quality while significantly suppressing noise. There are two main contributions:
\begin{itemize} 
    \item In the repairing stage, we introduce a causality-based knowledge distillation strategy to utilize future information in a causal way. Specifically, we employ a non-causal model as the teacher to guide a causal model in mimicking the behavior and predictions of the non-causal model. 

    \item In the denoising stage, the complex axial self-attention (ASA) along the frequency axis is introduced to overcome the limitation of the constraint receptive field. Self-attention can effectively capture long-range relations between features, leading the denoising network to extract more information from different frequencies.
\end{itemize}
Experimental results show that our approach achieves a 0.10 OVRL DNSMOS improvement on the ICASSP 2024 SSI Challenge blind test set compared to RaD-Net and outperforms the state-of-the-art system Gesper~\cite{gesper} on the ICASSP 2023 SSI Challenge blind test set.

\begin{figure*}[!htbp]
	\centering
	\begin{minipage}{1.0\linewidth}
		\centering
		\includegraphics[width=1.0\linewidth]{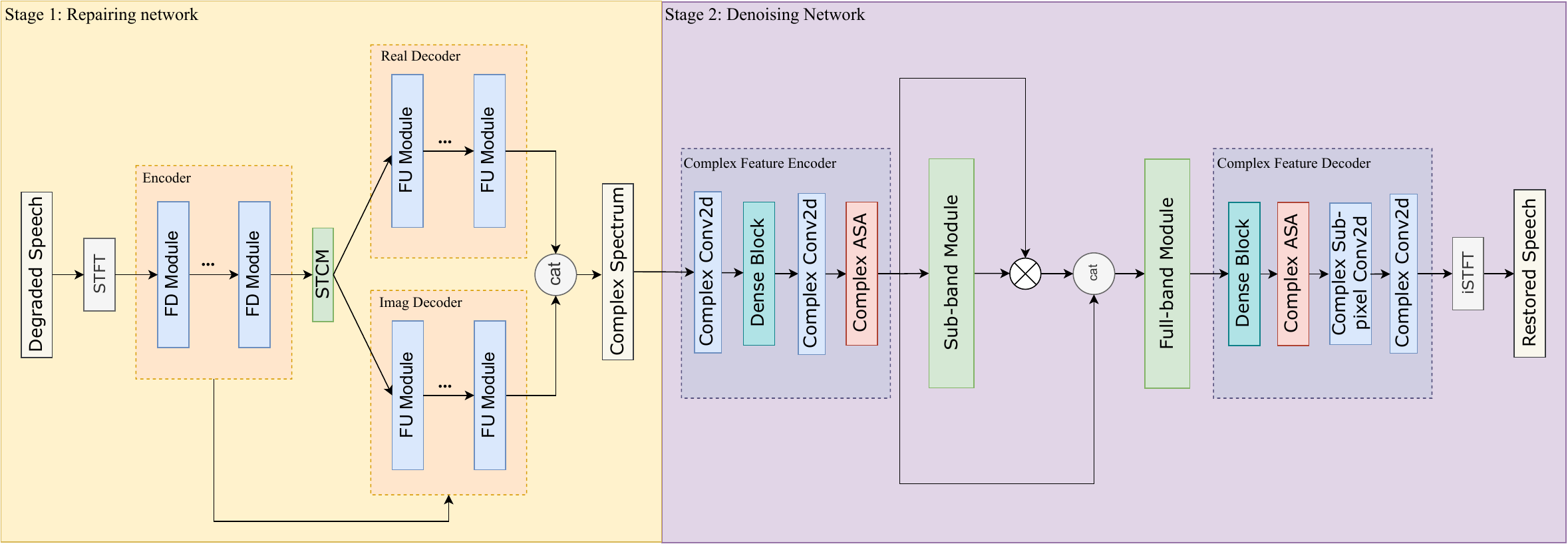}
		\label{f2_1}
	\end{minipage}
    \vspace{-0.8em}
    \caption{The architecture of RaD-Net 2.}
    \label{f2}
    \vspace{-0.7em}
\end{figure*}

\section{Proposed System}

As shown in Fig.~1, RaD-Net 2 follows the two-stage framework.  It is composed of two parts: the repairing network and the denoising network. We first perform the short-time Fourier transform (STFT) on the distorted speech to obtain the complex spectrum. Then the repairing network takes the degraded complex spectrum as input and outputs a relatively high-quality complex spectrum. Subsequently, the restored complex spectrum generated by the repairing network is fed to the denoising network, aiming to remove residual noise and artifacts. Finally, the inverse STFT (iSTFT) is applied to the output of the denoising network to obtain the final restored speech.


\subsection{Repairing Network}

Inspired by the superior performance of TEA-PSE~\cite{tea-pse}, we employ COM-Net from TEA-PSE as the backbone of the repairing network, which adopts an encoder-decoder architecture. Specifically, the encoder consists of three frequency down-sampling (FD) layers, while the decoders are stacked by three frequency up-sampling (FU) layers. The FD layer starts with a gated convolution layer (GateConv), followed by a cumulative layer norm, PReLU, and time-frequency convolution module (TFCM)~\cite{mtfaa}. The FU layer uses a mirror structure and replaces the GateConv with a transposed gated convolutional layer (TrGateConv). Between the encoder and decoder, we apply four stacked gated temporal convolutional modules (S-GTCM) for temporal modeling.

In the first stage, the repairing network restores multiple distortions simultaneously by generating the missing components of the degraded speech and preliminarily removing interfering signals such as noise. For generation tasks, the performance of models is influenced by their ability to access future information, resulting in non-causal models outperforming causal models.

To access future information and improve the performance of the causal repairing network, we introduce a causality-based knowledge distillation to use future information in a causal way. Specifically, the non-causal repairing network is employed as the teacher, and the causal repairing network is employed as the student. Convolution layers with non-causal padding and causal padding are adopted in the TFCM modules of the teacher and student model, respectively. Note that the teacher and the student model share the same parameter settings. The only difference is that the TFCM modules in the teacher model are non-causal while being causal in the student model. 

In the training stage, we first train the non-causal repairing network and freeze its parameters. Then, we pre-train the causal repairing network, which is applied to the student. Finally, we use the pre-trained non-causal repairing network as the teacher to perform knowledge distillation on the pre-trained causal repairing network. We adopt frequency-domain loss and adversarial loss to train both the teacher and student models. The frequency-domain loss consists of the spectral convergence loss $\mathcal{L}_{\text{sc}}$~\cite{scloss}, the L1 loss of the logarithmic magnitude $\mathcal{L}_{\text{log-mag}}$, and the asymmetric loss $\mathcal{L}_{\text{asym}}$, which can be defined as:
\begin{equation}
\begin{aligned}
    & \mathcal{L}_{\text{sc}}=\frac{||X-\hat{X}||_{F}}{||\hat{X}||_{F}}, \\
    & \mathcal{L}_{\text{log-mag}}=||log(X)-log(\hat{X})||_{1}, \\
    & \mathcal{L}_{\text{asym}}=||h(X^{0.5}-\hat{X}^{0.5})||_{2}^{2}, \\
    & h(x) = \left\{
        \begin{aligned}
            & 0, x\leq0 \\
            & x, x>0 \\
        \end{aligned}
        \right.
\end{aligned}
\end{equation}
where $X$ and $\hat{X}$ denote the magnitude spectrum of the clean speech and the restored speech during the training stage of the teacher model, while during the knowledge distillation stage, they represent the magnitude spectrum of the restored speech from the teacher model and the restored speech from the student model. The adversarial loss is composed of the generator loss $\mathcal{L}_{\text{G}}$~\cite{hifi} and the feature matching loss $\mathcal{L}_{\text{FM}}$~\cite{hifi}, which can be written as:
\begin{equation}
\begin{aligned}
    & \mathcal{L}_{\text{G}}=\mathbb{E}_{(\hat{s},s)}\Big[(D(s)-1)^2+(D(\hat{s}))^2\Big], \\
    & \mathcal{L}_{\text{FM}}=\mathbb{E}_{(\hat{s},s)}\Big[\sum_{i=1}^{M}\frac{1}{N_{i}}||D^{i}(s)-D^{i}(G(\hat{s}))||_1\Big]. \\
\end{aligned}
\end{equation}
where $s$ and $\hat{s}$ denote the clean speech and the restored speech in the training stage of the teacher model, while in the knowledge distillation stage, they represent the restored speech from the teacher model and the restored speech from the student model. M and $N_i$ denote the number of layers in each discriminator and the number of features in the i-th layer of the discriminator. The final loss can be represented as: 
\begin{equation}
\mathcal{L}_{\text{1}}=\mathcal{L}_{\text{sc}}+\mathcal{L}_{\text{log-mag}}+0.5\mathcal{L}_{\text{asym}}+\mathcal{L}_{\text{G}}+2\mathcal{L}_{\text{FM}}.
\end{equation}


\begin{figure}[htbp!]
	\begin{minipage}{1.0\linewidth}
		\centering
		\includegraphics[width=1.0\linewidth]{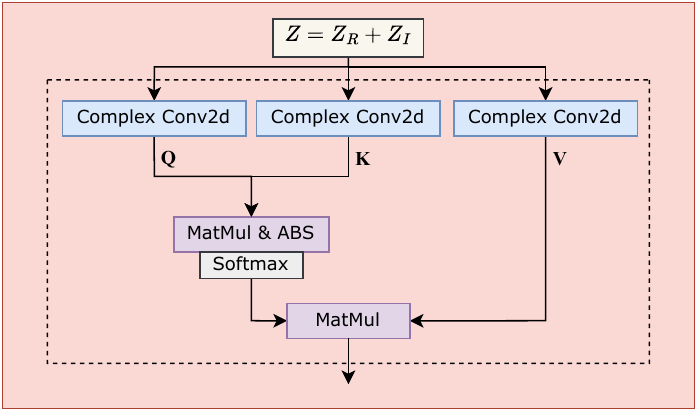}
		\label{f2_2}
	\end{minipage}
    \vspace{-0.8em}
    \caption{The architecture of the Complex ASA module.}
    \vspace{-0.6em}
    \label{f2}
\end{figure}

\subsection{Denoising Network}

In the second stage, a variant of S-DCCRN~\cite{s-dccrn} named S-DCCSN is applied as the denoising network. As shown in Fig.1, S-DCCSN is composed of four parts. The complex feature encoder (CFE) and complex feature decoder (CFD) extract additional information while maintaining low-frequency resolution. In addition, the sub-band and full-band modules are cascaded to process local and global frequency information, respectively. To reduce the computational cost, we replace convolution layers with depthwise separable convolution layers.

To overcome the limitation of the constraint receptive field, the complex ASA along the frequency axis is adapted in the second stage. Compared to pixel or patch-level attention in computer vision, the ASA~\cite{mtfaa} can improve the network’s ability to capture long-range relations between features while reducing the cost of memory and computation. As shown in Fig 1, we apply the complex ASA module in the CFE and CFD to effectively extract latent feature spaces from complex spectra. As shown in Fig.~2, the complex ASA module is composed of three complex convolution layers, which are used to obtain the queries $Q$, keys $K$, and values $V$:

\begin{small}
\begin{equation}
\begin{aligned}
    & \resizebox{.91\hsize}{!}{$ Q=(W_{R}^{Q}(Z_{R})-W_{I}^{Q}(Z_{I}))+j(W_{R}^{Q}(Z_{I})+W_{I}^{Q}(Z_{R})) ,$} \\
    & \resizebox{.91\hsize}{!}{$ K=(W_{R}^{K}(Z_{R})-W_{I}^{K}(Z_{I}))+j(W_{R}^{K}(Z_{I})+W_{I}^{K}(Z_{R})), $} \\
    & \resizebox{.91\hsize}{!}{$ V=(W_{R}^{V}(Z_{R})-W_{I}^{V}(Z_{I}))+j(W_{R}^{V}(Z_{I})+W_{I}^{V}(Z_{R})). $}
\end{aligned}
\end{equation}
\end{small}
where $W_{R}$ and $W_{I}$ denote the real and imaginary layers of a complex convolution layer $W$, respectively. The $Z_R$ and $Z_I$ are the real part and the imaginary part of the complex input. Inspired by the complex attention operation of D2Former~\cite{d2former}, the softmax function is employed on the absolute values of the complex matrix to compute the complex ASA. It can be defined as:
\begin{equation}
    \text{ComplexASA}(Q,K,V)=softmax(\frac{|QK^T|}{\sqrt{d_q}})V,
\end{equation}
where $d_p$ denotes the dimension of the queries $Q$. Note that $QK^T$ is the complex arithmetic operation, which is computed as:
\begin{equation}
    QK^T=(Q_{R}K_{R}^{T}-Q_{I}K_{I}^{T})+j(Q_{R}K_{I}^{T}+Q_{I}K_{R}^{}).
\end{equation}

As for training, we first pre-train the denoising network on denoising and dereverberation tasks without using discriminators. Subsequently, we load the pre-trained causal repairing and denoising networks and finetune only the parameters of the denoising network with discriminators. For the loss function, the asymmetric loss $\mathcal{L}_{\text{asym}}$, the scale-invariant signal-to-noise ratio (SI-SNR) loss $\mathcal{L}_{\text{si-snr}}$~\cite{conv-tasnet} and power-law compressed loss $\mathcal{L}_{\text{plc}}$ are applied in the pre-training stage. They can be computed as:
\begin{equation}
\begin{aligned}
    & \mathcal{L}_{\text{si-snr}}=20\log_{10}\frac{||(\hat{s}^{T}s/s^{T}\hat{s})\cdot s||}{||s-(\hat{s}^{T}s/s^{T}\hat{s})\cdot s||}, \\
    & \mathcal{L}_{\text{plc}}=||X^{0.5}e^{j\varphi X}-\hat{X}^{0.5}e^{j\varphi\hat{X}}||_{2}^{2}+||X^{0.5}-\hat{X}^{0.5}||_{2}^{2}. \\
\end{aligned}
\end{equation}
where $s$ and $\hat{s}$ denote the clean speech and the restored speech. $S$ and $\hat{S}$ are the complex spectra of the clean speech and the restored speech, respectively. Then we further incorporate $\mathcal{L}_{\text{G}}$ and $\mathcal{L}_{\text{FM}}$ to finetune the denoising network. The final loss used in the pre-train stage and fine-tune stage can be defined as:
\begin{equation}
\begin{aligned}
& \mathcal{L}_{\text{2-pre}}=\mathcal{L}_{\text{si-snr}}+\mathcal{L}_{\text{plc}}+\mathcal{L}_{\text{asym}}, \\
& \mathcal{L}_{\text{2}}=\mathcal{L}_{\text{si-snr}}+\mathcal{L}_{\text{plc}}+\mathcal{L}_{\text{asym}}+\mathcal{L}_{\text{G}}+2\mathcal{L}_{\text{FM}}. 
\end{aligned}
\end{equation}

\subsection{Discriminator}

Regarding the discriminators, we introduce multi-resolution discriminators (MRD)~\cite{disc} and multi-band discriminators (MBD)~\cite{mbd} in the training stage. 
The MRD is a mixture of 6 sub-discriminators, using magnitudes generated with different STFT parameters as input. Each sub-discriminator in MRD consists of 7 stacked convolution layers. Weight normalization and LeakyReLU are adopted after each convolution layer except for the last one. 
For MBD, complex spectra generated with different STFT parameters are divided into 5 subbands and then fed to 3 sub-discriminators. Each sub-discriminator applies 5 parallel convolutional modules to process sub-band spectra. The convolutional module consists of 5 stacked convolution layers, and weight normalization is adopted after each convolution layer.

\section{Experiments}

\subsection{Dataset}
We use a subset from the DNS5 dataset for training and evaluation. The subset includes approximately 400 hours of clean speech and 181 hours of noise clips sampled at 48kHz. In the ICASSP 2024 Speech Signal Improvement Challenge, a data synthesizer is released for data simulation. The data synthesizer generates degraded speech from clean speech by simulating the processing flow of real-world communication systems. After analyzing the blind test set and the simulation procedure of this synthesizer, several other data augmentation methods are introduced. Specifically, besides the GSM codec, we also introduce four other open-source codecs, namely OPUS, AAC, AMR-NB, and AMR-WB. We use the enhanced synthesizer to simulate a 1200-hour dataset for training and a 30-hour dataset for evaluation.


\begin{table*}[htbp]
\centering
\small
 \caption{DNSMOS and SIGMOS scores for the different approaches on the ICASSP 2024 SSI Challenge blind test set. The ``$\text{KD}_{\text{Large}}$" and ``$\text{KD}_{\text{Non-causal}}$" denote the knowledge distillation strategy using the large stage one model and the non-causal stage one model as the teacher, respectively. Among all the causal systems, the highest scores are highlighted below.}
\setlength{\tabcolsep}{3mm}
 \label{tab:dnsmos}
  \resizebox{\linewidth}{!}{ 
\begin{tabular}{@{}l|c|ccc|ccccccc@{}}
\toprule
    \multicolumn{1}{c|}{\multirow{2}{*}{Model}}     & \multicolumn{1}{c|}{\multirow{2}{*}{Para.~(M)}} & \multicolumn{3}{c|}{DNSMOS} & \multicolumn{7}{c}{SIGMOS} \\ 
     & & SIG & BAK & OVRL                                                                   & COL & DISC  & LOUD  & NOISE & REVERB & SIG & OVRL       \\ \midrule
    Noisy & - & 3.00 & 3.57 & 2.62                                                          & 3.09 & 3.89 & 3.55 & 3.44 & 3.70 & 3.05 & 2.54       \\ \midrule
    Stage-1 & 2.21 & 3.30 & 4.01 & 3.02                                                     & 3.53 & 4.01 & 4.07 & 4.25 & 4.21 & 3.45 & 3.03    \\
    Stage-1 Large & 3.54 & 3.34 & 4.05 & 3.08                                               & 3.58 & 4.04 & 4.05 & 4.31 & 4.25 & 3.49 & 3.09    \\
    Stage-1 Non-causal & 2.21 & 3.43 & 4.05 & 3.16                                          & 3.73 & 4.19 & 4.14 & 4.34 & 4.27 & 3.72 & 3.32    \\
    Stage-1 + $\text{KD}_{\text{Large}}$ & 2.21 & 3.31 & 4.01 & 3.04                                 & 3.55 & 4.03 & 4.06 & 4.27 & 4.22 & 3.47 & 3.06   \\
    Stage-1 + $\text{KD}_{\text{Non-causal}}$ & 2.21 & 3.36 & 4.02 & 3.09                            & \textbf{3.63} & 4.06 & 4.05 & 4.29 & 4.19 & 3.54 & 3.10    \\ \midrule
    RaD-Net & 4.00 & 3.34 & 4.10 & 3.10                                                     & 3.51 & 4.04 & 4.10 & 4.41 & 4.31 & 3.52 & 3.12    \\
    RaD-Net + ASA  & 3.97 & 3.36 & 4.13 & 3.14                                              & 3.51 & 4.05 & \textbf{4.12} & 4.48 & 4.36 & 3.59 & 3.17    \\
    RaD-Net + KD + ASA (RaD-Net 2) & 3.97 & \textbf{3.41} & \textbf{4.15} & \textbf{3.20}   & 3.62 & \textbf{4.09} & 4.11 & \textbf{4.56} & \textbf{4.42} & \textbf{3.69} & \textbf{3.27}  \\  
\bottomrule
\end{tabular}
}
\end{table*}

\subsection{Experimental Setup}

For all models used in experiments, we use a Hanning window with a 20ms frame length and a 10ms frame shift. The STFT length is 960, leading to 481-dim spectral features. To evaluate the effectiveness of our causality-based knowledge distillation strategy, we use the causal repairing network as the student, while using the large causal repairing network and the non-causal repairing network as the teacher. In the causal and non-causal repairing networks, the number of channels for all GConv and TrGConv are 64. The stride and kernel size of GConv and TrGConv are (4, 1), and (5, 1) in the frequency and time axis, respectively. One TFCM contains 3 depthwise dilated convolution layers with a dilation rate of \{1,2,4\} and a kernel size of (3, 5). One S-GTCM contains 4 GTCM layers with a kernel size of 5 for dilated convolution layers and a dilation rate of \{1,2,5,9\}, respectively. For the large causal repairing network, all GConv and TrGConv layers' channels are set to 80. One TFCM contains 4 depthwise dilated convolution layers with a dilation rate of \{1,2,4,8\}. In the denoising network, the number of channels for the sub-band module and the full-band module is \{16, 32, 32, 32, 64, 64\}, and the convolution kernel size and stride are set to (5,2) and (2,1) respectively. For the CED and CFD, the number of channels is 32 and the depth of DenseBlock is 5. For ASA, the number of hidden channels is 16. The hidden channels of STCM adopted by the sub-band module and full-band module in the denoising network are 64. Models are optimized by the AdamW~\cite{adamw} optimizer with an initial learning rate of 0.0002. The learning rate decay was scheduled by a 0.999 factor in every epoch.

\subsection{Metrics}
We use DNSMOS~\cite{dnsmos} and SIGMOS~\cite{sig2024} to evaluate different enhancement systems. The DNSMOS is a non-intrusive perceptual objective metric to evaluate wide-band speech in three dimensions, namely speech quality (SIG), noise suppression (BAK), and overall quality (OVRL). The SIGMOS is a new non-intrusive objective metric released in the ICASSP 2024 SSI Challenge. It is based on the ITU-T P.804 standard and designed for full-band speech. The SIGMOS evaluates speech quality in seven dimensions: speech signal (SIG), coloration (COL), discontinuity (DISC), loudness (LOUD), noisiness (NOISE), Reverb (REVERB), and overall quality (OVRL). 
\section{Results And Analysis}





We conduct ablation experiments to evaluate the effectiveness of our proposed approaches. From Table 1, it can be observed that the repairing network using the non-causal model as the teacher outperforms the repairing network without performing knowledge distillation, which indicates that the non-causal repairing network effectively improves the performance of the causal repairing network. In addition, we adjust the parameter settings to ensure that the denoising network with ASA and the denoising network without ASA maintain approximately the same computational complexity. In this case, the denoising network with the complex ASA achieves improvements on SIG, BAK, and OVRL compared to the denoising network without the complex ASA.

SIGMOS scores for different systems in Table 1 exhibit a similar trend, further demonstrating the aforementioned conclusions. Moreover, upon analyzing each dimension of SIGMOS, we can draw the following conclusions: 1) The repairing network can improve each dimension in SIGMOS of the degraded speech, while the denoising network mainly improves NOISE and REVERB.  This trend matches our previous analysis that the repairing network performs main speech restoration and preliminary noise suppression. On the other hand, the denoising network removes residual noise and artifacts. 2) The non-causal repairing network achieves the best performance in SIGMOS, while not performing as well on DNSMOS. We analyze this result from two aspects: metric evaluation and types of data distortions. The DNSMOS and SIGMOS are designed to evaluate wide-band speech and full-band speech. In the blind test set, some speech utterances missed high-frequency components. Therefore, the SIGMOS evaluates the quality of generated high-frequency components that DNSMOS can not access. The above analysis indicates that future information is a significant factor in the generation task. 3) The student model, which uses the non-causal model as the teacher,  achieves improvements in nearly all dimensions of SIGMOS compared to the student model, which uses the larger causal model as the teacher. This improvement indicates that the non-causal repairing network transforms more knowledge into the student model.


We further evaluate RaD-Net 2 on the ICASSP 2023 SSI Challenge blind test set. As shown in Table 3, our proposed system outperforms Gesper\cite{gesper}, which ranked 1st in the ICASSP 2023 SSI Challenge.
\label{ssec:subhead}
\begin{table}[htbp]
\centering
\small
 \caption{DNSMOS on ICASSP 2023 SSI Challenge blind set.}
\setlength{\tabcolsep}{7mm}
 \label{tab:dnsmos}
  \resizebox{\linewidth}{!}{
\begin{tabular}{@{}lccc@{}}
\toprule
    Model       & SIG & BAK & OVRL \\ \midrule
    Noisy      & 2.89 & 3.45 & 2.46    \\
    Gesper      & 3.45 & 4.12 & 3.20    \\
    RaD-Net 2 & \textbf{3.47} & \textbf{4.13} & \textbf{3.22}    \\
\bottomrule
\end{tabular}
}
\end{table}

\vspace{-1.0em}
\section{Conclusions}

In this paper, we extend the two-stage neural network RaD-Net to RaD-Net 2. In the first stage, a causality-based knowledge distillation strategy is introduced, using the non-causal repairing network to use future information in a causal way. Then in the second stage, we introduce the complex axial self-attention module into the CFE and CFD to improve the denoising network's ability to capture long-range relations between different frequencies.



\bibliographystyle{IEEEtran}
\bibliography{mybib}

\end{document}